\documentclass[aps,pra,superscriptaddress,twocolumn,showpacs]{revtex4}%
\usepackage{graphics}
\usepackage{amsmath}
\usepackage{graphicx}
\usepackage{amsfonts}
\usepackage{amssymb}

\begin{document}
\title{Quantum reading under a local energy constraint}
\author{Gaetana Spedalieri}
\email{gae.spedalieri@york.ac.uk}
\affiliation{Department of Computer Science, University of York, York YO10 5GH, United Kingdom}
\author{Cosmo Lupo}
\author{Stefano Mancini}
\affiliation{School of Science and Technology, University of Camerino, Camerino 62032, Italy}
\author{Samuel L. Braunstein}
\author{Stefano Pirandola}
\affiliation{Department of Computer Science, University of York, York YO10 5GH, United Kingdom}

\begin{abstract}
Nonclassical states of light play a central role in many quantum information
protocols. Very recently, their quantum features have been exploited to
improve the readout of information from digital memories, modeled as arrays of
microscopic beam splitters [S. Pirandola, Phys. Rev. Lett. \textbf{106},
090504 (2011)]. In this model of \textquotedblleft quantum
reading\textquotedblright, a nonclassical source of light with
Einstein-Podolski-Rosen correlations has been proven to retrieve more
information than any classical source. In particular, the quantum-classical
comparison has been performed under a global energy constraint, i.e., by
fixing the mean total number of photons irradiated over each memory cell. In
this paper we provide an alternative analysis which is based on a local energy
constraint, meaning that we fix the mean number of photons \textit{per signal
mode} irradiated over the memory cell. Under this assumption, we investigate
the critical number of signal modes after which a nonclassical source of light
is able to beat any classical source irradiating the same number of signals.

\end{abstract}

\pacs{03.67.--a, 03.65.Ud, 42.50.--p, 89.20.Ff, 89.70.Cf}
\maketitle

\section{Introduction}

Quantum information has disclosed a modern approach to both quantum mechanics
and information theory \cite{NielsenBook}. Very recently, this field has been
further developed into the so-called \textquotedblleft continuous
variable\textquotedblright\ domain, where information is encoded and processed
by using quantum systems with infinite-dimensional Hilbert spaces
\cite{BraREV,BraREV2,GaussSTATES,GaussSTATES2} (see also the recent
review~\cite{RMP}). The most important example of these systems are the
bosonic modes of the electromagnetic field, today manipulated with very high
precision in quantum optics labs. Thus, in the continuous variable framework,
a wide range of results have been successfully achieved, including protocols
of quantum
teleportation~\cite{CVtelepo,Bra98,RalphTELE,PirTeleOPMecc,Barlett2003,Sherson}%
, teleportation networks~\cite{TeleNET,teleREV,teleJMO,Pirgames,Pir2005},
entanglement swapping~\cite{Entswap,Entswap2,PirENTswap}, quantum
cryptography~\cite{QKD0,QKD1,Weed,Weed2,CharacATT,Chris,PirNATURE,Devetak,PirSKcapacity,QKDreview}%
, quantum
computation~\cite{Qcomp1,Qcomp2,Qcomp2b,Qcomp2c,Qcomp2d,Qcomp2e,Qcomp2f,QcREF,QcREF2}
and cluster quantum
computation~\cite{Qcomp3,Qcomp5,Qcomp6,Qcomp7,clusterREF,clusterREF2}.

One of the key resources in quantum information is quantum entanglement. In
the bosonic setting, quantum entanglement is usually present under the form of
Einstein-Podolski-Rosen (EPR) correlations~\cite{EPRpaper}, where the
quadrature operators of two separate bosonic modes are so correlated to beat
the standard quantum limit~\cite{NOTEshot}. The simplest source of EPR
correlations is the two-mode squeezed vacuum (TMSV) state. In the number-ket
representation, this state is defined by
\begin{equation}
\left\vert \xi\right\rangle =(\cosh\xi)^{-1}\sum_{n=0}^{\infty}(\tanh\xi
)^{n}\left\vert n\right\rangle _{s}\left\vert n\right\rangle _{i}~,
\end{equation}
where $\xi$ is the squeezing parameter and $\{s,i\}$ is an arbitrary pair of
bosonic modes, that we may call \textquotedblleft signal\textquotedblright%
\ and \textquotedblleft idler\textquotedblright. In particular, $\xi$
quantifies the signal-idler entanglement and determines the mean number of
photons \textrm{sinh}$^{2}\xi$ in each mode. Since it is entangled, the TMSV
state cannot be prepared by applying local operations and classical
communications (LOCCs) to a couple of vacua $\left\vert 0\right\rangle
_{s}\otimes\left\vert 0\right\rangle _{i}$ or to any other kind of tensor
product state. For this reason, the TMSV state cannot be expressed as a
classical mixture of coherent states $\left\vert \alpha\right\rangle
_{s}\otimes\left\vert \beta\right\rangle _{i}$ with $\alpha$ and $\beta$
arbitrary complex amplitudes. In other words, its
P-representation~\cite{Prepres,Prepres2}%
\begin{equation}
\left\vert \xi\right\rangle \left\langle \xi\right\vert =\int\int d^{2}\alpha
d^{2}\beta\boldsymbol{~}\mathcal{P}(\alpha,\beta)~\left\vert \alpha
\right\rangle _{s}\left\langle \alpha\right\vert \otimes\left\vert
\beta\right\rangle _{i}\left\langle \beta\right\vert ~,
\end{equation}
involves a function $\mathcal{P}$ which is non-positive and, therefore, cannot
be considered as a genuine probability distribution. For this reason, the
TMSV\ state is a particular kind of \textquotedblleft
nonclassical\textquotedblright\ state. Other kinds are single-mode squeezed
states and Fock states. By contrast a bosonic state is called
\textquotedblleft classical\textquotedblright\ when its P-representation is
positive, meaning that the state can be written as a classical mixture of
coherent states. Thus a classical source of light is composed by a set of $m$
bosonic modes in a state%
\begin{equation}
\rho=\int d^{2}\alpha_{1}\cdots\int d^{2}\alpha_{m}\boldsymbol{~}%
\mathcal{P}(\alpha_{1},\cdots,\alpha_{m})~\otimes_{k=1}^{m}\left\vert
\alpha_{k}\right\rangle \left\langle \alpha_{k}\right\vert ~, \label{Prepres}%
\end{equation}
where $\mathcal{P}$ is positive and normalized to $1$. Typically, classical
sources are just made by a collection of coherent states with amplitudes
$\{\bar{\alpha}_{1},\cdots,\bar{\alpha}_{m}\}$, i.e., $\rho=\otimes_{k=1}%
^{m}\left\vert \bar{\alpha}_{k}\right\rangle \left\langle \bar{\alpha}%
_{k}\right\vert $ which corresponds to having
\begin{equation}
\mathcal{P}=\prod_{k=1}^{m}\delta^{2}(\alpha_{k}-\bar{\alpha}_{k})~.
\end{equation}
In other situations, where the sources are particularly chaotic, they are
better described by a collection of thermal states with mean photon numbers
$\{\bar{n}_{1},\cdots,\bar{n}_{m}\}$, so that%
\begin{equation}
\mathcal{P}=\prod_{k=1}^{m}\frac{\exp(-\left\vert \alpha_{k}\right\vert
^{2}\bar{n}_{k})}{\pi\bar{n}_{k}}~.
\end{equation}
More generally, we can have classical states which are not just tensor
products but they have (classical) correlations among different bosonic modes.

The comparison between classical and nonclassical states has clearly triggered
a lot of interest. The main idea is to compare the use of a candidate
nonclassical state, like the EPR state, with all the classical states for
specific information tasks. One of these tasks is the detection of
low-reflectivity objects in far target regions under the condition of
extremely low signal-to-noise ratios. This scenario has been called
\textquotedblleft quantum illumination\textquotedblright\ and has been
investigated in a series of papers~\cite{QIll1,QIll2,QIll3,Guha,Devi,YuenNair}.

More recently, EPR correlations have been exploited for a completely different
task in a completely different regime of parameters. In the model of
\textquotedblleft quantum reading\textquotedblright\ \cite{QreadingPRL}, EPR
correlations have been used to retrieve information from digital memories
which are reminiscent of today's optical disks, such as CDs and DVDs. A
digital memory can in fact be modelled as a sequence of cells corresponding to
beam splitters with two possible reflectivities $r_{0}$ and $r_{1}$ (used to
encode a bit of information). By fixing the mean total number of photons
$N$\ irradiated over each memory cell, it is possible to show that a
non-classical source of light with EPR correlations retrieves more information
than any classical source~\cite{QreadingPRL}. In general, the improvement is
found in the regime of few photons ($N=1-100$)\ and for memories with high
reflectivities, as typical for optical memories. In this regime, the gain of
information given by quantum reading can be dramatic, i.e., close to $1$ bit
for each bit of the memory. Further studies on quantum reading of memories
have been pursued by several
authors~\cite{Nair11,Hirota11,QreadCAP,Bisio11,Arno11,Saikat,Saikat2}. In
particular, Ref.~\cite{Nair11} has shown that other non-classical states, such
as Fock states, can have remarkable advantages over classical sources.
Ref.~\cite{Hirota11} has proposed an alternative model of quantum reading
based on a binary phase encoding. Ref.~\cite{Bisio11} has further studied the
problem of binary discrimination in optical devices. Ref.~\cite{Arno11} has
both proposed and experimentally implemented a model of unambiguous quantum
reading. Ref.~\cite{QreadCAP} has defined the notion of quantum reading
capacity, a quantity which has been also investigated in Ref.~\cite{Saikat2}.
Finally, Ref.~\cite{Saikat} has proposed explicit capacity-achieving receivers
for quantum reading.

\begin{figure}[ptbh]
\vspace{-0.1cm}
\par
\begin{center}
\includegraphics[width=0.5\textwidth] {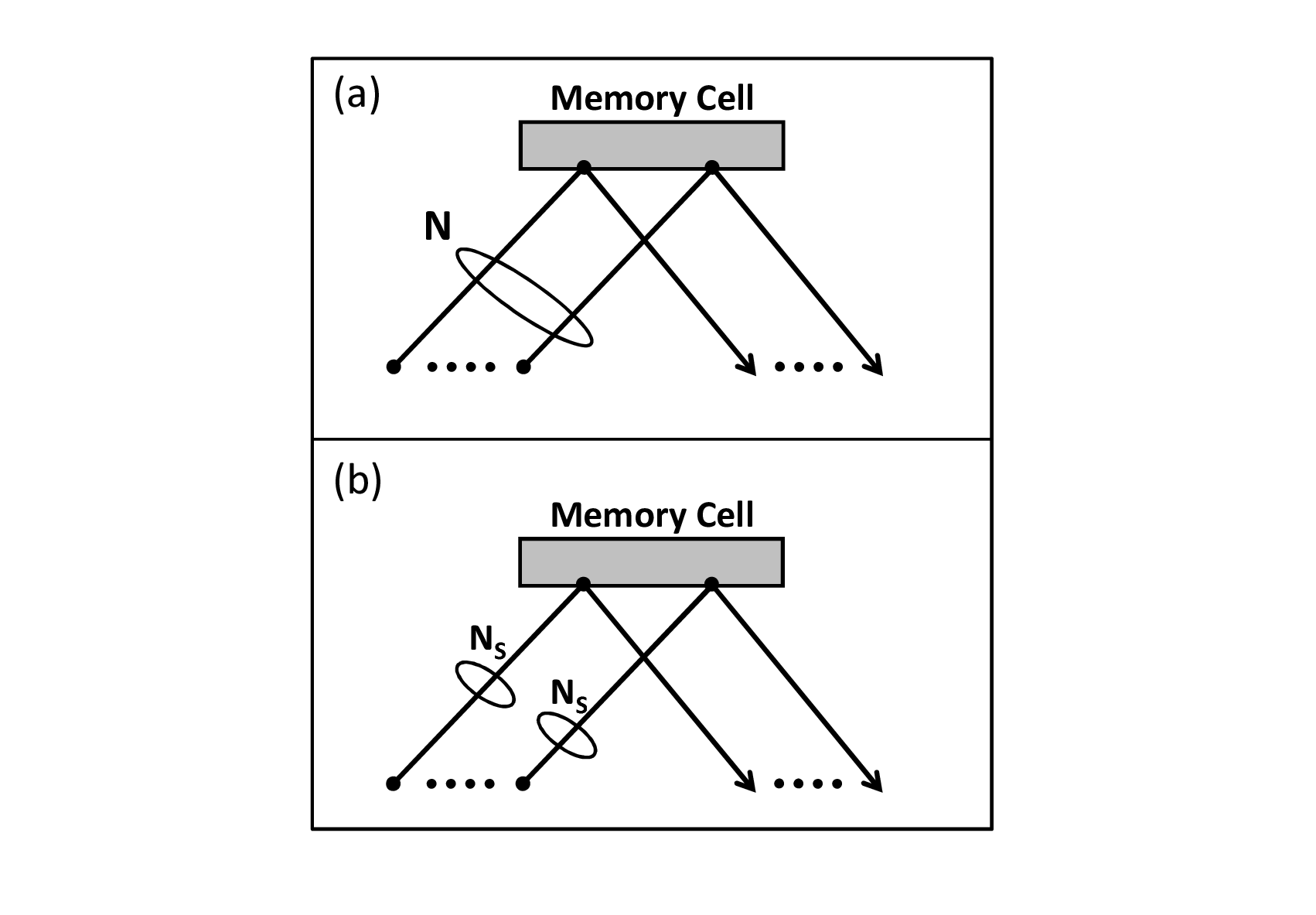}
\end{center}
\par
\vspace{-0.9cm}\caption{Inset (a). Quantum reading of Ref.~\cite{QreadingPRL}
is formulated under a global energy constraint. This means that we fix the
total average number of photons $N$ irradiated over the memory cell. Thus, if
the number of input signals is $M$, each one has an average of $N/M$ photons,
which goes to zero for $M\rightarrow\infty$. Inset (b). In this paper, we
consider an alternative model of quantum reading under a local energy
constraint (locally-constrained quantum reading). In this case, we fix the
average number of photons $N_{S}$ for each input signal. Since the total
number of signals $M$ can be arbitrary, we have that the total energy
irradiated over the cell $MN_{S}$ is generally unbounded.}%
\label{energyPIC}%
\end{figure}

It is fundamental to remark that an important point in the study of
Ref.~\cite{QreadingPRL} is that the quantum-classical comparison is performed
under a \textit{global energy constraint}, i.e., by fixing the total average
number of photons $N$\ which are irradiated over each memory cell [see
Fig.~\ref{energyPIC}(a)]. Under this assumption, it is possible to construct
an EPR transmitter, made by a suitable number of TMSV states, which is able to
outperform \textit{any} classical source composed by \textit{any} number of
modes. 

In this paper we consider a different kind of comparison: we fix the number of
signal modes irradiated over the target cell ($M$) and the mean number of
photons \textit{per signal mode} ($N_{S}$). Under these assumptions, we
compare an EPR transmitter with a classical source. Then, for fixed $N_{S}$,
we determine the critical number of signal modes $M^{(N_{S})}$ after which an
EPR\ transmitter with $M>$ $M^{(N_{S})}$ is able to beat any classical source
(with the same number of signals $M$). Since we are here fixing the average
number of photons per signal mode, our energy constraint is now \textit{local}%
:\ it restricts the energy of each signal mode but not the energy of the total
set of signal modes. We call this alternative model \textquotedblleft
locally-constrained quantum reading\textquotedblright\ [see
Fig.~\ref{energyPIC}(b)].

The difference between global and local energy constraints is also discussed
in Ref.~\cite{RMP} for the general problem of Gaussian channel discrimination.
Mathematically speaking, both these energy constraints make the problem of
channel discrimination non-trivial in the continuous variable setting, where
the use of infinite energy always allows one to distinguish two Gaussian
channels in a perfect way. In the presence of a global energy constraint, the
error probability in the Gaussian channel discrimination is generally
different from zero and the problem is to find the minimum value. In the
presence of a local energy constraint, the error probability goes to zero with
the number $M$ of signals and the general problem is to study its convergence,
i.e., finding the best error exponent~\cite{RMP}. From this point of view, the
present paper shows that the best convergence of the error probability has to
be found within the set of non-classical states.

From a practical point of view, the use of a local energy constraint is useful
in all those situations where the energy of each radiation mode has to be
taken under control. For instance, consider a photosensitive organic memory
where data is encoded in error correcting blocks. For simplicity, we may think
of blocks of $M$ cells where information is encoded by means of an $M$-bit
repetition code (the generalization to more complex codes such as the
Reed-Solomon codes is only a technical issue~\cite{Cover}). In this scenario,
the stored information can be safely retrieved from the block if we irradiate
a single mode per cell with suitable low energy (for instance, a single
temporal mode, i.e., a pulse, with a mean energy $N_{S}$ which is below the
critical energy associated with the photodegradation of the material). By
contrast, optimizing the readout under a global energy constraint may be
unsafe in this specific situation, since the optimal readout of the block
could be achieved by concentrating all the available energy into a single
mode. Thus, if we use a total of $N=MN_{S}$ mean photons, we could have all
these photons irradiated over a single cell of the block, with inevitable
damage for the memory.

The remainder of the paper is structured as follows. In Sec.~\ref{Sec1} we
review the basic readout mechanism of quantum reading specifying the analysis
to the case of a local energy constraint. Then, in Sec.~\ref{Sec2}, we
explicitly show how EPR\ correlations can be used to beat any classical source
of light in the readout of information. Finally, Sec.~\ref{Sec3} is for
conclusions. Note that we also provide two appendices. In
Appendix~\ref{AppMETHODS} we discuss the general mathematical methods used in
our derivations, and Appendix~\ref{AppTECH} contains some technical proofs.

\section{Readout mechanism\label{Sec1}}

\begin{figure}[ptbh]
\vspace{-0.6cm}
\par
\begin{center}
\includegraphics[width=0.5\textwidth] {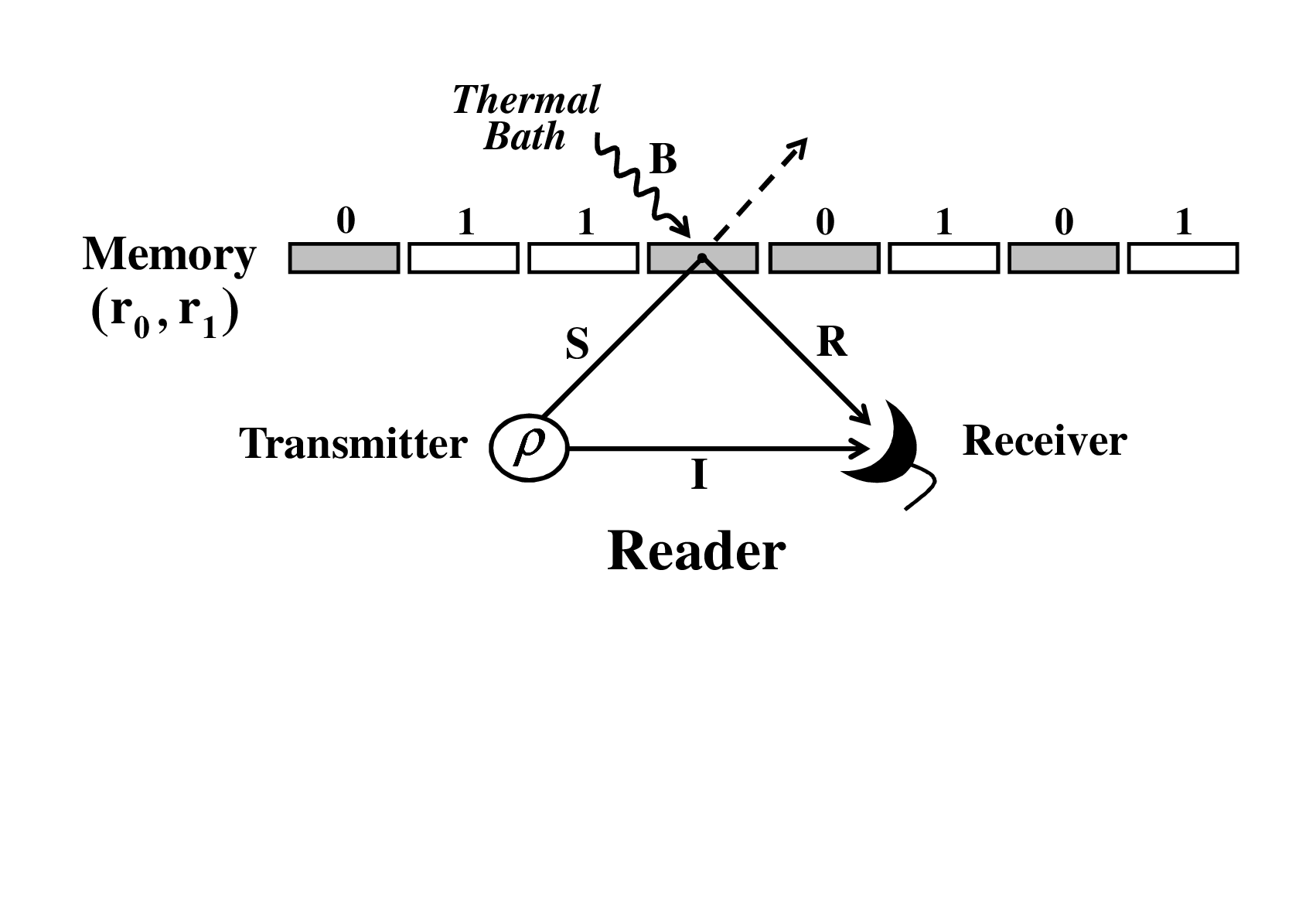}
\end{center}
\par
\vspace{-2.7cm}\caption{\textbf{Model of memory}. Digital information is
stored in a disk whose memory cells are beam splitter mirrors with different
reflectivities: $r=r_{0}$ encoding bit-value $u=0$ and $r=r_{1}$ encoding
bit-value $u=1$. \textbf{Readout}. A reader is generally composed by a
transmitter and a receiver. It retrieves a stored bit by probing a memory cell
with a signal system $S$ (composed of $M$ bosonic modes) and detecting the
reflected system $R$ together with an idler system $I$ (composed of $L$
bosonic modes). In general, the output system $R$ combines the signal system
$S$ with a bath system $B$ ($M$ bosonic modes in thermal states). The
transmitter is in a state $\rho$ which can be classical (a classical
transmitter) or non-classical (a quantum transmitter). In particular, we
consider a quantum transmitter with EPR\ correlations between the signal and
idler systems. In this paper, the quantum-classical comparison is performed
under a local energy constraint, i.e., by fixing the average number of photons
$N_{S}$ per signal mode (the signal system $S$ has a total average number of
photons $N=MN_{S}$ which is generally unbounded).}%
\label{QreadPIC}%
\end{figure}

Here we briefly review the basic readout mechanism of Ref.~\cite{QreadingPRL},
specifying the study to the case of a local energy constraint. Consider a
model of a digital optical memory (or disk) where the memory cells are beam
splitter mirrors with different reflectivities $r=r_{0},r_{1}$ (with
$r_{1}\geq r_{0}$). In particular, the bit-value $u=0$ is encoded in a
lower-reflectivity mirror ($r=r_{0}$), that we may call a \textit{pit}, while
the bit-value $u=1$ is encoded in a higher-reflectivity mirror ($r=r_{1}$),
that we may call a \textit{land} (see\ Fig.~\ref{QreadPIC}). Close to the
disk, a reader aims to retrieve the value of the bit $u$ which is stored in
each memory cell. For this purpose, the reader exploits a transmitter (to
probe a target cell) and a receiver (to measure the corresponding output). In
general, the transmitter consists of two quantum systems, called
\textit{signal} $S$ and \textit{idler} $I$, respectively. The signal system
$S$ is a set of $M$ bosonic modes which are directly shined on the target
cell. The mean total number of photons of this system is simply given by
$N=MN_{S}$, where $N_{S}$ is the mean number of photons per signal mode
(simply called \textquotedblleft energy\textquotedblright, hereinbelow). At
the output of the cell, the reflected system $R$ is combined with the idler
system $I$, which is a supplementary set of bosonic modes whose number $L$ can
be completely arbitrary. Both the systems $R$ and $I$ are finally measured by
the receiver (see\ Fig.~\ref{QreadPIC}).

We assume that Alice's apparatus is very close to the disk, so
that no significant source of noise is present in the gap between the disk and
the decoder. However, we assume that non-negligible noise comes from the
thermal bath present at the other side of the disk. This bath generally
describes stray photons, transmitted by previous cells and bouncing back to
hit the next ones. For this reason, the reflected system $R$ combines the
signal system $S$ with a bath system $B$ of $M$ modes. These environmental
modes are assumed in a tensor product of thermal states, each one with $N_{B}$
mean photons (white thermal noise). Thus, in this model we identify five basic
parameters: the reflectivities of the memory $\{r_{0},r_{1}\}$, the
temperature of the bath $N_{B}$, and the profile of the signal $\{M,N_{S}\}$,
which is given by the number of signals $M$\ and the energy $N_{S}$.

In general, for a fixed input state $\rho$ at the transmitter (systems $S,I$),
Alice will get two possible output states $\sigma_{0}$ and $\sigma_{1}$ at the
receiver (systems $R,I$). These output states are the effect of two different
quantum channels, $\mathcal{E}_{0}$ and $\mathcal{E}_{1}$, which depend on the
bit $u=0,1$ stored in the target cell. In particular, we have
\begin{equation}
\sigma_{u}=(\mathcal{E}_{u}\otimes\mathcal{I})(\rho)~,
\end{equation}
where the conditional channel $\mathcal{E}_{u}$ acts on the signal system,
while the identity channel $\mathcal{I}$ acts on the idler system. More
precisely, we have $\mathcal{E}_{u}=\mathcal{R}_{u}^{\otimes M}$, where
$\mathcal{R}_{u}$ is a one-mode lossy channel with conditional loss $r_{u}$
and fixed thermal noise $N_{B}$. Now, the minimum error probability $P_{err}$
affecting the decoding of $u$ is just the error probability affecting the
statistical discrimination of the two output states, $\sigma_{0}$ and
$\sigma_{1}$, via an optimal receiver. This quantity is equal to
\begin{equation}
P_{err}=[1-D(\sigma_{0},\sigma_{1})]/2~,
\end{equation}
where $D(\sigma_{0},\sigma_{1})$ is the trace distance between $\sigma_{0}$
and $\sigma_{1}$~\cite{Helstrom,Fuchs,FuchsThesis}. Clearly, the value of
$P_{err}$ determines the average amount of information which is decoded for
each bit stored in the memory. This quantity is equal to
\begin{equation}
J=1-H(P_{err})~,
\end{equation}
where
\begin{equation}
H(x):=-x\log_{2}x-(1-x)\log_{2}(1-x)
\end{equation}
is the usual formula for the binary Shannon entropy. In the following, we
compare the performance of decoding in two paradigmatic situations, one where
the transmitter is described by a non-classical state (a quantum transmitter)
and one where the transmitter is in a classical state (a classical
transmitter). In particular, we show how a quantum transmitter with EPR
correlations (an EPR transmitter) is able to outperform classical
transmitters. The quantum-classical comparison is performed for a fixed signal
profile $\{M,N_{S}\}$. Then, for various fixed values of the energy $N_{S}$
(local energy constraint), we study the critical number of signal modes
$M^{(N_{S})}$ after which an EPR transmitter (with $M>M^{(N_{S})}$ signals) is
able to beat any classical transmitter (with the same number of signals $M$).

\section{Quantum-classical comparison\label{Sec2}}

First let us consider a classical transmitter. A classical transmitter with
$M$ signals and $L$ idlers is described by a classical state $\rho$ as
specified by Eq.~(\ref{Prepres}) with $m=M+L$. In other words it is a
probabilistic mixture of multi-mode coherent states $\otimes_{k=1}%
^{M+L}\left\vert \alpha_{k}\right\rangle \left\langle \alpha_{k}\right\vert $.
Given this transmitter, we consider the corresponding error probability
$P_{err}^{class}$ which affects the readout of the memory. Remarkably, this
error probability is lower-bounded by a quantity which depends on the signal
profile $\{M,N_{S}\}$, but not from the number $L$ of the idlers and the
explicit expression of the $\mathcal{P}$-function. In fact, we
have~\cite{QreadingPRL}%
\begin{equation}
P_{err}^{class}\geq\mathcal{C}(M,N_{S}):=\frac{1-\sqrt{1-F(N_{S})^{M}}}{2}~,
\label{CB_cread}%
\end{equation}
where $F(N_{S})$ is the fidelity between $\mathcal{R}_{0}(|N_{S}^{1/2}%
\rangle\langle N_{S}^{1/2}|)$ and $\mathcal{R}_{1}(|N_{S}^{1/2}\rangle\langle
N_{S}^{1/2}|)$, the two possible outputs of the single-mode coherent state
$|N_{S}^{1/2}\rangle\langle N_{S}^{1/2}|$ (see Appendix~\ref{AppMETHODS} for
more details). As a consequence, all the classical transmitters with signal
profile $\{M,N_{S}\}$ retrieve an information which is upper-bounded by
\begin{equation}
J_{class}:=1-H[\mathcal{C}(M,N_{S})]~.
\end{equation}

Now, let us construct a transmitter having the same signal profile
$\{M,N_{S}\}$, but possessing EPR correlations between signals and idlers.
This is realized by taking $M$ identical copies of a TMSV state, i.e.,
$\rho=\left\vert \xi\right\rangle \left\langle \xi\right\vert ^{\otimes M}$
where $N_{S}=\mathrm{sinh}^{2}\xi$. Given this transmitter, we consider the
corresponding error probability $P_{err}^{quant}$ affecting the readout of the
memory. This quantity is upper-bounded by the quantum Chernoff bound
\cite{QCbound,QCbound2,QCbound3,QCbound4,MinkoPRA}%
\begin{equation}
P_{err}^{quant}\leq\mathcal{Q}(M,N_{S}):=\frac{1}{2}\left[  Q(N_{S})\right]
^{M}, \label{QCB_qread}%
\end{equation}
where%
\begin{equation}
Q(N_{S}):=\inf_{s\in(0,1)}\mathrm{Tr}(\theta_{0}^{s}\theta_{1}^{1-s})~,
\label{QCB_qread2}%
\end{equation}
and
\begin{equation}
\theta_{u}:=(\mathcal{R}_{u}\otimes\mathcal{I})(\left\vert \xi\right\rangle
\left\langle \xi\right\vert )~.
\end{equation}
Since $\theta_{0}$ and $\theta_{1}$ are Gaussian states, we can write out
their symplectic decompositions~\cite{Alex} and compute the quantum Chernoff
bound using the formula for multimode Gaussian states given in
Ref.~\cite{MinkoPRA} (see Appendix~\ref{AppMETHODS} for more details). Then,
we can easily compute a lower bound
\begin{equation}
J_{quant}:=1-H[\mathcal{Q}(M,N_{S})]
\end{equation}
for the information which is decoded via this quantum transmitter.

In order to show an improvement with respect to the classical case, it is
sufficient to prove the positivity of the \textquotedblleft information
gain\textquotedblright\
\begin{equation}
G:=J_{quant}-J_{class}~.
\end{equation}
This quantity is in fact a lower bound for the average information which is
gained by using the EPR quantum transmitter over any classical transmitter.
Roughly speaking, the value of $G$ estimates the minimum information which is
gained by the quantum readout for each bit of the memory. In general, $G$ is a
function of all the basic parameters of the model, i.e., $G=G(M,N_{S}%
,r_{0},r_{1},N_{B})$. Numerically, we can easily find signal profiles
$\{M,N_{S}\}$, classical memories $\{r_{0},r_{1}\}$, and thermal baths $N_{B}%
$, for which we have the quantum effect $G>0$. Some of these values are
reported in the following table.%
\[%
\begin{tabular}
[c]{|c|c|c|c|c|c|}\hline
$~M~$ & $~~N_{S}~~$ & $~~~~r_{0}~~~~$ & $~~~~r_{1}~~~~$ & $~~N_{B}~~$ &
$~~~G~($bits$)~~~$\\\hline
$1$ & $3.5$ & $0.5$ & $0.95$ & $0.01$ & $~6.2\times10^{-3}$\\\hline
$10$ & $1$ & $0.2$ & $0.8$ & $0.01$ & $~3.4\times10^{-2}$\\\hline
$30$ & $1$ & $0.38$ & $0.85$ & $1$ & $~1.2\times10^{-3}$\\\hline
$100$ & $0.1$ & $0.25$ & $0.85$ & $0.01$ & $~5.9\times10^{-2}$\\\hline
$200$ & $0.1$ & $0.6$ & $0.95$ & $0.01$ & $0.22$\\\hline
$2\times10^{5}$ & $0.01$ & $0.995$ & $1$ & $0$ & $0.99$\\\hline
\end{tabular}
\ \ \
\]
Note that we can find choices of parameters where $G\simeq1$, i.e., the
classical readout of the memory does not decode any information whereas the
quantum readout is able to retrieve all of it. As shown in the last row of the
table, this situation can occur when both the reflectivities of the memory are
very close to $1$. From the first row of the table, we can observe another
remarkable fact: for a land-reflectivity $r_{1}$\ sufficiently close to $1$,
one signal with few photons can give a positive gain. In other words, the use
of a single, but sufficiently entangled, TMSV state $\left\vert \xi
\right\rangle \left\langle \xi\right\vert $ can outperform any classical
transmitter, which uses one signal mode with the same energy (and potentially
infinite idler modes).

Here an importat point to remark is that, once that we find a positive gain
$G>0$, this positivity is preserved if we increase the number of signals $M$.
In other words, if $G$ is positive for some $\tilde{M}$, then it is positive
for every $M\geq\tilde{M}$ (keeping the other parameters fixed.) This is
trivial to prove. In fact, $G(\tilde{M})>0$ is equivalent to $\mathcal{Q}%
(\tilde{M},N_{S})<\mathcal{C}(\tilde{M},N_{S})$ which is equivalent to
\begin{equation}
\frac{1}{2}Q^{\tilde{M}}<\frac{1-\sqrt{1-F^{\tilde{M}}}}{2}~,
\end{equation}
according to Eqs.~(\ref{CB_cread}) and~(\ref{QCB_qread}). This means that
\begin{equation}
Q<\left(  1-\sqrt{1-F^{\tilde{M}}}\right)  ^{1/\tilde{M}}~.
\end{equation}
For every $M\geq\tilde{M}$, we then have%
\begin{equation}
Q^{M}<(1-\sqrt{1-x})^{m}~,
\end{equation}
where $m:=M/\tilde{M}$ and $x:=F^{\tilde{M}}$. Now, we can use the algebraic
inequality%
\begin{equation}
(1-\sqrt{1-x})^{m}\leq1-\sqrt{1-x^{m}}~,
\end{equation}
which holds for every $m\geq1$ and $x\in\lbrack0,1]$. Then, we get%
\begin{equation}
Q^{M}<1-\sqrt{1-F^{M}}~,
\end{equation}
which is equivalent to%
\begin{equation}
\mathcal{Q}(M,N_{S})<\mathcal{C}(M,N_{S})~,
\end{equation}
for every $M\geq\tilde{M}$.

Thanks to this property, for given reflectivities $\{r_{0},r_{1}\}$ and bath
temperature $N_{B}$, i.e., for a fixed memory, if a quantum transmitter with
signal profile $\{\tilde{M},N_{S}\}$ outperforms the classical transmitters,
then any other quantum transmitter with the same energy $N_{S}$ and
$M\geq\tilde{M}$ is also able to beat the classical readout.

It is also important to note that the advantage of quantum transmitters over
classical transmitters is asymptotically negligible in the limit of large
number of signals. Mathematically speaking, the information gain
$G=G(M,N_{S},r_{0},r_{1},N_{B})$ always goes to zero for $M\rightarrow+\infty
$. This is clearly a consequence of the specific constraint that we consider
in this work, for which the limit of $M\rightarrow+\infty$ corresponds to the
limit of infinite energy, a regime where any transmitter is able to retrieve
information with negigible error probability. In fact, given a memory with two
reflectivities $r_{0}\neq r_{1}$ and finite temperature $N_{B}$, an arbitrary
transmitter in any tensor product state $\rho=\omega^{\otimes M}$ has error
probability%
\begin{equation}
P_{err}\leq\frac{Q^{M}}{2}~,
\end{equation}
where the quantum Chernoff bound $Q$ is evaluated over the single-copy ouput
states. Now, for non-zero signal energy $N_{S}>0$, we have $Q<1$, so that
$P_{err}\rightarrow0$ for $M\rightarrow+\infty$. The situation is clearly
different from Ref.~\cite{QreadingPRL}, where the global energy contraint is
adopted, i.e., the mean total number of photons $N$ is fixed. In that case,
the broadband limit $M\rightarrow+\infty$ implies a vanishing energy per
signal mode $N_{S}=NM^{-1}$. As a result, we have $Q\rightarrow1$ and the
upperbound does no longer guarantee that $P_{err}$\ tends to zero. As a matter
of fact, in Ref.~\cite{QreadingPRL}, the broadband limit is absolutely
nontrivial and gives the optimal gain for the most important class of memories.

Contrarily to what happens in Ref.~\cite{QreadingPRL}, in the present model of
locally-constrained quantum reading we have that the maximum advantage, i.e.,
the optimal gain $G$, occurs for intermediate values of the signal mode number
$M$. Given a memory with parameters $\{r_{0},r_{1},N_{B}\}$, there is an
optimal range of numbers $M$ depending on the signal energy of the transmitter
$N_{S}$. In order to numerically investigate this behavior, we consider an
estimate for the information gain $G^{\ast}\leq G$ which is provided by using
the quantum Battacharyya bound in the place of the quantum Chernoff
bound~\cite{MinkoPRA}. In other words, we consider%
\begin{equation}
G^{\ast}:=J_{quant}^{\ast}-J_{class}~,
\end{equation}
where%
\begin{equation}
J_{quant}^{\ast}:=1-H[\mathcal{B}(M,N_{S})]
\end{equation}
and%
\begin{equation}
\mathcal{B}(M,N_{S}):=\frac{1}{2}\left[  \mathrm{Tr}(\theta_{0}^{1/2}%
\theta_{1}^{1/2})\right]  ^{M}%
\end{equation}
is the quantum Battacharyya bound computed over the two equiprobable output
states $\theta_{0}$ and $\theta_{1}$.

\begin{figure}[ptbh]
\vspace{-0.1cm}
\par
\begin{center}
\includegraphics[width=0.48\textwidth] {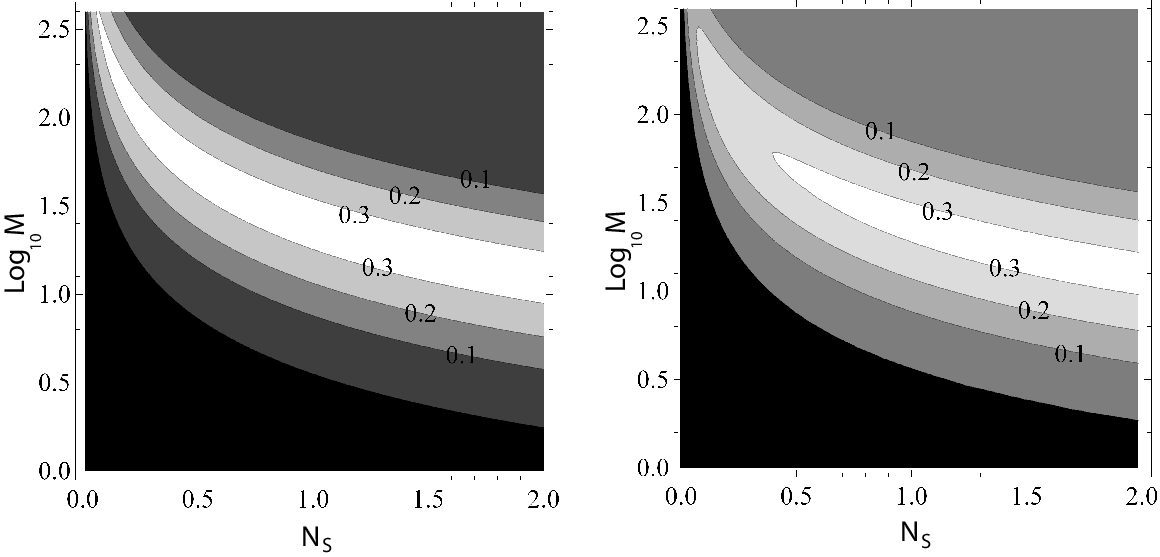}
\end{center}
\par
\vspace{-0.6cm}\caption{Contourplots of the information gain $G^{\ast}$ as a
function of the signal energy $N_{S}$ and the number of signal modes $M$ (in
logarithmic scale). Reflectivities are $r_{0}=0.6$ and $r_{1}=0.95$. Thermal
noise is $N_{B}=10^{-3}$ (left plot) and $N_{B}=10^{-2}$ (right plot). In the
bottom black area, we have $G^{\ast}=0$. The maximum values of $G^{\ast}$ are
taken in the intermediate white area where $G^{\ast}\gtrsim0.3$ bits. For
large number of modes $M$, we have $G^{\ast}\rightarrow0$.}%
\label{FIGab}%
\end{figure}

\begin{figure}[ptbh]
\vspace{-0.1cm}
\par
\begin{center}
\includegraphics[width=0.48\textwidth] {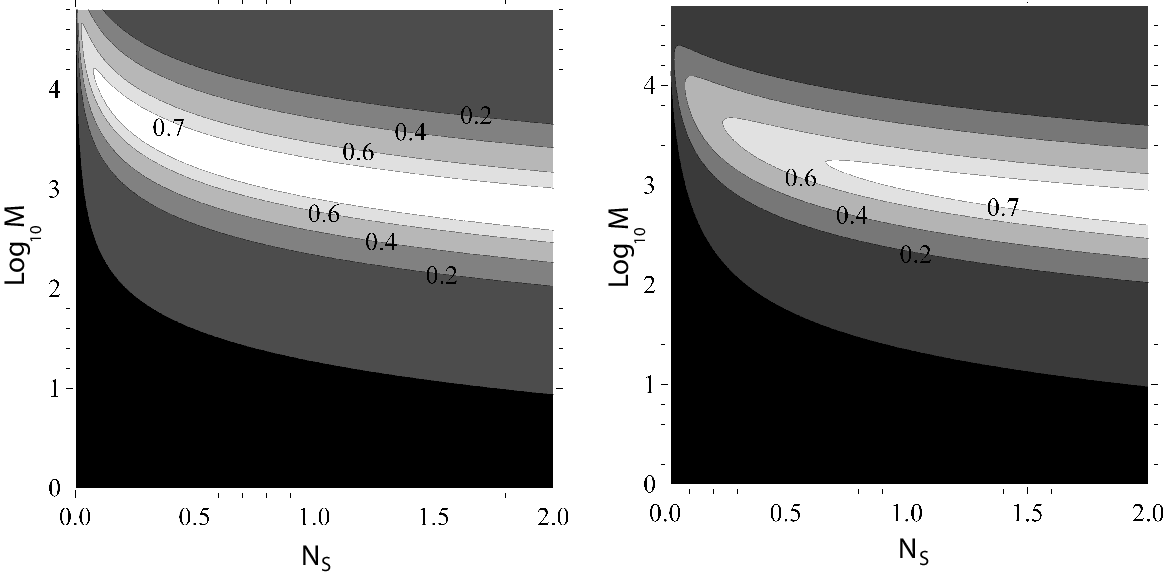}
\end{center}
\par
\vspace{-0.6cm}\caption{Contourplots of the information gain $G^{\ast}$ as a
function of the signal energy $N_{S}$ and the number of signal modes $M$ (in
logarithmic scale). Reflectivities are $r_{0}=0.95$ and $r_{1}=0.98$. Thermal
noise is $N_{B}=10^{-3}$ (left plot) and $N_{B}=10^{-2}$ (right plot). In the
bottom black area, we have $G^{\ast}=0$. The maximum values of $G^{\ast}$ are
taken in the intermediate white area where $G^{\ast}\gtrsim0.7$~bits. For
large number of modes $M$, we have $G^{\ast}\rightarrow0$.}%
\label{FIGcd}%
\end{figure}

\begin{figure}[ptbh]
\vspace{-0.1cm}
\par
\begin{center}
\includegraphics[width=0.48\textwidth] {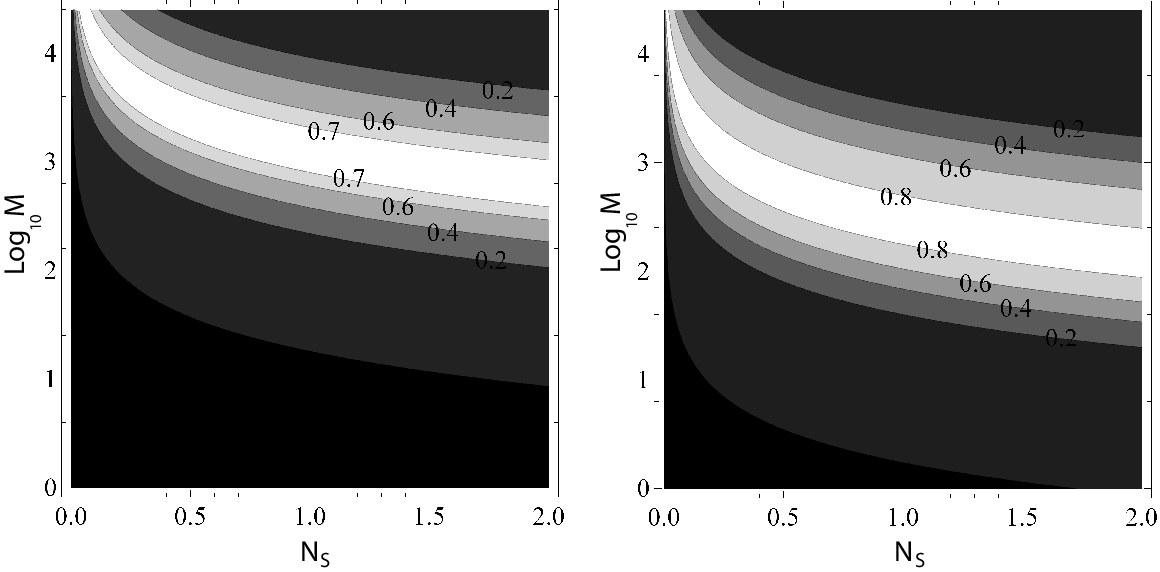}
\end{center}
\par
\vspace{-0.6cm}\caption{Contourplots of the information gain $G^{\ast}$ as a
function of the signal energy $N_{S}$ and the number of signal modes $M$ (in
logarithmic scale). Thermal noise is equal to $N_{B}=10^{-5}$. In the left
plot, reflectivities are $r_{0}=0.95$ and $r_{1}=0.98$. In the right plot,
reflectivities are $r_{0}=0.95$ and $r_{1}=0.999$. In the bottom black area,
we have $G^{\ast}=0$. The maximum values of $G^{\ast}$ are taken in the
intermediate white area where $G^{\ast}\gtrsim0.7$ (left) and $G^{\ast}%
\gtrsim0.8$ (right). For large number of modes $M$, we have $G^{\ast
}\rightarrow0$.}%
\label{FIGef}%
\end{figure}

Given a memory specified by a set of parameters $\{r_{0},r_{1},N_{B}\}$, we
can study the information gain $G^{\ast}$ as a function of the signal profile
$\{M,N_{S}\}$. This is done in Figs.~\ref{FIGab}, \ref{FIGcd}, and~\ref{FIGef}%
. As we can see from Fig.~\ref{FIGab}, the information gain $G^{\ast}$ is zero
for low values of $M$ (black area in the figure). It takes its maximum for $M$
belonging to an intermediate range of values (this range corresponds to the
white area in the figure). Then, for higher values of $M$, the value of
$G^{\ast}$ gradually decreases to zero. For a memory with reflectivities
$r_{0}=0.6$ and $r_{1}=0.95$ and affected by a thermal noise $N_{B}%
=10^{-3}-10^{-2}$, we can reach an optimal gain $G^{\ast}\gtrsim0.3$ by using
around $M=10$ modes with $N_{S}=2$. At lower energies, we achieve the same
performance by using more signal modes.

In Fig.~\ref{FIGcd}, we consider a memory of better quality, i.e., with higher
reflectivities (equal to $r_{0}=0.95$ and $r_{1}=0.98$, respectively). As we
can see from the figure, the information gain can reach optimal values above
$0.7$~bits if we consider around $M=10^{3}$ signals.

Finally, in Fig.~\ref{FIGef}, we consider even better memories, with high
reflectivities and low thermal noise. As we can see from the figure, gains
above $0.8$~bits can be reached by using $M=10^{2}-10^{3}$ signal
modes.

\subsection{Ideal memories}

\begin{figure}[ptbh]
\vspace{+0.0cm}
\par
\begin{center}
\includegraphics[width=0.38\textwidth] {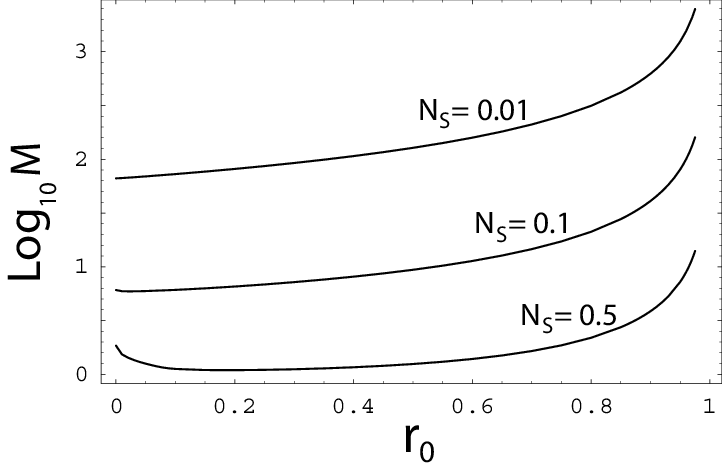}
\end{center}
\par
\vspace{-0.5cm}\caption{Number of signals $M$ (logarithmic scale)\ versus
pit-reflectivity $r_{0}$. The curves refer to $N_{S}=0.01$, $0.1$ and $0.5$
photons. For each value of the energy $N_{S}$, we plot the critical number
$M^{(N_{S})}(r_{0})$ as function of $r_{0}$. All the curves have an asymptote
at $r_{0}=1$. For $N_{S}\gtrsim2.5$ photons (curves not shown), we have
another asymptote at $r_{0}=0$. }%
\label{PRLmin}%
\end{figure}

According to our numerical investigation, the quantum readout is generally
more powerful when the land-reflectivity is sufficiently high (i.e.,
$r_{1}\gtrsim0.8$). For this reason, it is very important to analyze the
scenario in the limit of ideal land-reflectivity ($r_{1}=1$). Using the
terminology of Ref.~\cite{QreadingPRL}, we call an \textquotedblleft ideal
memory\textquotedblright\ a classical memory with $r_{1}=1$. Clearly, this
memory is completely characterized by the value of its pit-reflectivity
$r_{0}$. For ideal memories, the quantum Chernoff bound of
Eq.~(\ref{QCB_qread}) takes an analytical form given by the \textquotedblleft
Chernoff term\textquotedblright%
\begin{equation}
Q(N_{S})=\frac{1}{[1+(1-\sqrt{r_{0}})N_{S}]^{2}+N_{B}(2N_{S}+1)(1-r_{0})},
\end{equation}
and the classical bound of Eq.~(\ref{CB_cread}) can be computed using
\begin{equation}
F(N_{S})=\gamma^{-1}\exp[-\gamma^{-1}(1-\sqrt{r_{0}})^{2}N_{S}]~,
\label{fidTEXT}%
\end{equation}
where $\gamma:=1+(1-r_{0})N_{B}$ (see Appendix~\ref{AppMETHODS} for more
details). Using these formulas, we can study the behavior of the gain $G$ in
terms of the remaining parameters $\{M,N_{S},r_{0},N_{B}\}$. Let us consider
an ideal memory with generic $r_{0}\in\lbrack0,1)$ in a generic thermal bath
$N_{B}\geq0$. For a fixed energy $N_{S}$, we consider the minimum number of
signals $M^{(N_{S})}$ above which $G>0$. To be precise, the critical number
$M^{(N_{S})}$ that we consider is a solution of the equation $G=0$. From this
real value we derive the minimum number of signals (which is an integer) by
taking its ceiling function $\lceil M^{(N_{S})}\rceil$. The critical number
$M^{(N_{S})}$ can be defined independently from the thermal noise $N_{B}$ by
performing a numerical maximization over $N_{B}$. Then, for a given value of
the energy $N_{S}$, the critical number $M^{(N_{S})}$ becomes a function of
$r_{0}$ alone, i.e., $M^{(N_{S})}=M^{(N_{S})}(r_{0})$. Its behavior is shown
in Fig.~\ref{PRLmin} for different values of the energy. 

It is remarkable that, for low-energy signals ($N_{S}=0.01-1$ photons), the
critical number $M^{(N_{S})}(r_{0})$ is finite for every $r_{0}\in\lbrack
0,1)$. This means that, for ideal memories and low-energy signals, there
always exists a finite number of signals $M^{(N_{S})}$ above which the quantum
readout of the memory is more efficient than its classical readout. In other
words, there is an EPR\ transmitter with $M>M^{(N_{S})}$ able to beat any
classical transmitter with the same number of signals $M$. In the low-energy
regime considered, $M^{(N_{S})}(r_{0})$ is relatively small for almost all the
values of $r_{0}$, except for $r_{0}\rightarrow1$\ where $M^{(N_{S})}%
(r_{0})\rightarrow\infty$. In fact, for $r_{0}\simeq1$, we derive
\begin{equation}
M^{(N_{S})}(r_{0})\simeq\lbrack4N_{S}(2N_{S}+1)(1-r_{0})]^{-1}~,
\end{equation}
which diverges at $r_{0}=1$. Such a divergence is expected, since we must have
$P_{err}^{quant}=P_{err}^{class}=1/2$\ for $r_{0}=r_{1}$ (see
Appendix~\ref{AppTECH} for details). Apart from the divergence at $r_{0}=1$,
in all the other points $r_{0}\in\lbrack0,1)$, the critical number
$M^{(N_{S})}(r_{0})$ decreases for increasing energy $N_{S}$ (see
Fig.~\ref{PRLmin}). In particular, for $N_{S}=1$ photon, we have $M^{(N_{S}%
)}(r_{0})\simeq1$\ for most of the reflectivities $r_{0}$. In other words, for
energies around one photon, a single TMSV state is sufficient to provide a
positive gain for most of the ideal memories. However, the decreasing trend of
$M^{(N_{S})}(r_{0})$ does not continue for higher energies ($N_{S}\geq1$). In
fact, just after $N_{S}=1$, $M^{(N_{S})}(r_{0})$ starts to increase around
$r_{0}=0$. In particular, for $N_{S}\geq1$, we can derive
\begin{equation}
M^{(N_{S})}(0)\simeq(\ln2)[2\ln(1+N_{S})-N_{S}]^{-1}~,
\end{equation}
which is increasing in $N_{S}$, and becomes infinite at $N_{S}\simeq2.5$. As a
consequence, for $N_{S}\gtrsim2.5$ photons, we have a second asymptote
appearing at $r_{0}=0$ (see Appendix~\ref{AppTECH} for more details). This
means that the use of high-energy signals ($N_{S}\gtrsim2.5$) does not assure
positive gains for memories with extremal reflectivities $r_{0}=0$ and
$r_{1}=1$.

\section{Conclusion\label{Sec3}}

In conclusion, we have considered the basic model of digital memory studied in
Ref.~\cite{QreadingPRL}, which is composed of beam splitter mirrors with
different reflectivities. Adopting this model, we have compared an EPR
transmitter with classical sources for fixed signal profiles, finding positive
information gains for memories with high land-reflectivities ($r_{1}%
\gtrsim0.8$). Analytical results can be derived in the limit of ideal
land-reflectivity ($r_{1}=1$) which defines the regime of ideal memories. In
this case, by fixing the mean number of photons per signal mode (local energy
constraint), we have computed the critical number of signals above which an
EPR\ transmitter gives positive information gains, therefore beating any
classical transmitter. For low-energy signals ($0.01-1$ photons) this critical
number is finite and relatively small for every ideal memory. In particular,
an EPR\ transmitter with one TMSV state can be sufficient to achieve positive
information gains for almost all the ideal memories.

Thus our results corroborate the outcomes of Ref.~\cite{QreadingPRL} providing
an alternative study which considers a local energy constraint instead of a
global one. As discussed in Ref.~\cite{QreadingPRL} and its supplementary
materials, potential applications are in the technology of optical digital
memories where we could increase data-transfer rates and storage capacities.
For instance, let us fix the mean signal power $P$ which is irradiated on the
memory cell during the readout time $t$. This is approximately given by
$P=h\nu Nt^{-1}$, where $h$ is the Planck constant, $\nu$\ is the carrier
frequency, and $N$ is mean total number of photons. Suppose that we can access
low values of $N_{S}$ using a reasonable low number of modes $M$, so that the
value of $N$ is globally low. Now, at fixed power and frequency, the
low-energy regime (low $N$) corresponds to short readout times $t$, i.e., high
data transfer rates. Equivalently, at fixed power and readout time, the
low-energy regime corresponds to high frequencies $\nu$, i.e., dense storage devices.

Finally, another potential application is the readout of organic digital
memories, which are devices extremely photosensitive at high frequencies. In
this case, the use of faint quantum signals could safely read the data without
damaging the storing devices. As discussed before, locally-constrained quantum
reading may be particularly suitable for the readout of these fragile memories
thanks to the direct control of the mean energy of each radiation mode.

\appendix

\section{Methods\label{AppMETHODS}}

Here we provide more details about the methods used in our computations. We
start with a brief review of the bosonic Gaussian states
(Appendix~\ref{AppGAUSS}). Then, we give a detailed description of the readout
problem discussing the main techniques for computing the performances of
classical and nonclassical transmitters (Appendix~\ref{BasicSECTION}).
Finally, we consider the special case of ideal memories, for which we can
derive simple analytical formulas (Appendix~\ref{AppIDEAL}).

\subsection{Bosonic systems and Gaussian states\label{AppGAUSS}}

A bosonic system is generally composed of $n$ modes. This means that the
system is associated to a tensor product Hilbert space $\mathcal{H}^{\otimes
n}$ and described by a vector of quadrature operators
\begin{equation}
\mathbf{\hat{x}}^{T}:=(\hat{q}_{1},\hat{p}_{1},\ldots,\hat{q}_{n},\hat{p}%
_{n})~,
\end{equation}
which satisfy the commutation relations $[\hat{x}_{k},\hat{x}_{l}%
]=2i\mathbf{\Omega}_{kl}$, where%
\begin{equation}
\mathbf{\Omega}:=\bigoplus\limits_{i=1}^{n}\left(
\begin{array}
[c]{cc}%
0 & 1\\
-1 & 0
\end{array}
\right)  \label{Symplectic_Form}%
\end{equation}
defines a symplectic form (correspondingly, a real matrix $\mathbf{S}$ is
called \textquotedblleft symplectic\textquotedblright\ if $\mathbf{S\Omega
S}^{T}=\mathbf{\Omega}$).

By definition, a bosonic state $\rho$ is \textquotedblleft
Gaussian\textquotedblright\ if its Wigner function is Gaussian~\cite{RMP}. As
a result, a Gaussian state $\rho$ is fully characterized by its first and
second order statistical moments. These are the displacement vector
\begin{equation}
\mathbf{\bar{x}}:=\mathrm{Tr}(\mathbf{\hat{x}}\rho)~,
\end{equation}
and the covariance matrix (CM) $\mathbf{V}$, with generic element%
\begin{equation}
V_{kl}:=\tfrac{1}{2}\mathrm{Tr}\left(  \{\hat{x}_{k},\hat{x}_{l}\}\rho\right)
-\bar{x}_{k}\bar{x}_{l}~,
\end{equation}
where $\{,\}$ is the anticommutator. The CM is a $2n\times2n$ real symmetric
matrix which must satisfy the uncertainty principle~\cite{SIMONprinc,RMP}%
\begin{equation}
\mathbf{V}+i\mathbf{\Omega}\geq0~. \label{unc_PRINC}%
\end{equation}
According to Williamson's theorem~\cite{Willy}, every CM\ $\mathbf{V}$ can be
decomposed in the form%
\begin{equation}
\mathbf{V}=\mathbf{SWS}^{T}~,
\end{equation}
where $\mathbf{S}$ is a symplectic matrix and%
\begin{equation}
\mathbf{W}=\bigoplus\limits_{i=1}^{n}\nu_{i}\mathbf{I}=\left(
\begin{array}
[c]{ccccc}%
\nu_{1} &  &  &  & \\
& \nu_{1} &  &  & \\
&  & \ddots &  & \\
&  &  & \nu_{n} & \\
&  &  &  & \nu_{n}%
\end{array}
\right)
\end{equation}
is called the \textquotedblleft Williamson form\textquotedblright\ of
$\mathbf{V}$. In this matrix, the diagonal elements $\{\nu_{1},\cdots,\nu
_{n}\}$ represent the \textquotedblleft symplectic spectrum\textquotedblright%
\ of $\mathbf{V}$. The symplectic spectrum provides powerful ways to express
physical properties of the Gaussian state. For instance, the uncertainty
principle can be formulated as~\cite{Alex,RMP}%
\begin{equation}
\mathbf{V}>0~,~\nu_{i}\geq1~.
\end{equation}

\subsection{Quantum reading versus classical reading\label{BasicSECTION}}

In our model of classical memory, each memory cell is represented by a beam
splitter mirror with two possible reflectivities, i.e., the pit-reflectivity
$r_{0}$ and the land-reflectivity $r_{1}$. This dichotomic choice $r\in
\{r_{0},r_{1}\}$ is used to encode an information bit $u\in\{0,1\}$ in the
memory cell. Then, one side of the memory is subject to decoding, while the
other side is affected by white thermal noise, with average photon number per
mode equal to $N_{B}$. It is clear that this model of memory corresponds to a
problem of Gaussian channel discrimination. In fact, each memory cell can be
seen as an attenuator channel, transforming an input signal mode into an
output reflected mode. In particular, this attenuator channel has a
transmission efficiency (or \textquotedblleft linear loss\textquotedblright)
which is given by the dichotomic reflectivity of the cell $r\in\{r_{0}%
,r_{1}\}$, and a thermal noise which is fixed and equal to $N_{B}$. Depending
on the bit $u\in\{0,1\}$ which is stored in the cell, we then have two
possible attenuator channels, that we denote by $\mathcal{R}_{0}$\ and
$\mathcal{R}_{1}$. In other words, the unknown bit $u$ stored in the cell is
encoded into a conditional attenuator channel $\mathcal{R}_{u}$.

Let us describe explicitly the action of $\mathcal{R}_{u}$. Given the
quadratures $\mathbf{\hat{x}}_{s}^{T}:=(\hat{q}_{s},\hat{p}_{s})$ of an input
signal mode $s$, the quadratures $\mathbf{\hat{x}}_{r}$ of the output
reflected mode $r$ are given by the Heisenberg relation%
\begin{equation}
\mathbf{\hat{x}}_{r}=\sqrt{r}\mathbf{\hat{x}}_{s}+\sqrt{1-r}\mathbf{\hat{x}%
}_{b}~, \label{couplingSB}%
\end{equation}
where $r\in\{r_{0},r_{1}\}$ and $\mathbf{\hat{x}}_{b}$ are the quadratures of
a bath mode $b$. In particular, the bath mode is described by a thermal state
$\rho_{b}(N_{B})$ with $N_{B}$ average photons, i.e., a Gaussian state with
zero mean and CM%
\begin{equation}
\mathbf{V}_{b}=(2N_{B}+1)\mathbf{I~.}%
\end{equation}

Once that we have specified the action of a memory cell over an arbitrary
signal mode $s$, we can analyze its full action on an arbitrary transmitter.
In general, we have a system $S$\ of $M$ signal modes impinging on the cell,
besides an ancillary system $I$ of $L$ idler modes which bypass the cell. At
the output of the cell, the system $R$ of the $M$ reflected modes is combined
with the idler system $I$ in a joint measurement at the receiver. The
fundamental parameters of the transmitter are contained in its signal profile
$\{M,N_{S}\}$, which is composed by the number of signal modes $M$ and the
average number of photons per signal $N_{S}$.

Let us denote by $\rho_{SI}$\ the global state of the input systems $\{S,I\}$.
The memory cell does not affect the idler system $I$, but acts on the signal
system $S$ by coupling every signal mode $s\in S$ with an independent thermal
mode $b$, which belongs to a bath system $B$ in the multimode thermal state
\begin{equation}
\rho_{B}=\rho_{b}(N_{B})^{\otimes M}~.
\end{equation}
Since the action on the signal system $S$ is one-mode and conditional, the
global state of the output systems $\{R,I\}$ can be written as%
\begin{equation}
\rho_{RI}(u)=\left(  \mathcal{R}_{u}^{\otimes M}\otimes\mathcal{I}^{\otimes
L}\right)  (\rho_{SI})~,
\end{equation}
where $\mathcal{I}^{\otimes L}$ is the identity channel acting on the idler
system. For a fixed state $\rho_{SI}$ at the transmitter, we have a
conditional output state $\rho_{RI}(u)$ at the receiver, which depends on the
bit $u$ stored in the memory cell. Thus, the minimum error probability in
decoding the stored bit is just the error probability affecting the optimal
discrimination of the two output states $\rho_{RI}(0)$ and $\rho_{RI}(1)$. As
we know, this error probability is equal to~\cite{Helstrom}%
\begin{equation}
P_{err}=\frac{1}{2}\left\{  1-D[\rho_{RI}(0),\rho_{RI}(1)]\right\}  ~,
\end{equation}
where $D[\rho_{RI}(0),\rho_{RI}(1)]$ is the trace distance between $\rho
_{RI}(0)$ and $\rho_{RI}(1)$. Clearly, the value of $P_{err}$ determines the
average amount of information which is decoded for each bit stored in the
memory. This average information is equal to $J=1-H(P_{err})$, where $H(x)$ is
the usual formula of the binary Shannon entropy.

In our work, we estimate the average decoded information $J$ in two paradigmic
situations, i.e., for a quantum transmitter with EPR correlations ($J_{Q}$),
and for a generic classical transmitter ($J_{C}$). By fixing the signal
profile $\{M,N_{S}\}$, we compare $J_{Q}$\ and $J_{C}$. More exactly, we fix
all the basic parameters of the model, i.e., besides fixing the signal profile
$\{M,N_{S}\}$, we also fix the reflectivities of the memory $\{r_{0},r_{1}\}$
and the thermal noise $N_{B}$. Then, we investigate what are the values of the
basic parameters $\{M,N_{S},r_{0},r_{1},N_{B}\}$ for which $J_{Q}>J_{C}$. In
particular, for proving this enhancement, we compare a lower bound of $J_{Q}$
with an upper bound of $J_{C}$.

\subsubsection{Classical transmitters}

Let us start considering an arbitrary classical transmitter. For a classical
transmitter with $M$ signals and $L$ idlers we can exploit the classical
discrimination bound proven in Ref.~\cite{QreadingPRL}. The minimum error
probability $P_{err}^{class}$ affecting the readout of the memory cell is
lower-bounded by $\mathcal{C}(M,N_{S})$ in Eq.~(\ref{CB_cread}) where
$F(N_{S})$ is the fidelity between the two states $\mathcal{R}_{0}%
(|N_{S}^{1/2}\rangle\langle N_{S}^{1/2}|)$ and $\mathcal{R}_{1}(|N_{S}%
^{1/2}\rangle\langle N_{S}^{1/2}|)$. Using the formula of the fidelity for
single-mode Gaussian states~\cite{fidG2,fidG3,fidG4}, we get%
\begin{equation}
F(N_{S})=\frac{1}{\sqrt{\gamma^{2}+\theta}-\sqrt{\theta}}\exp\left[
-\frac{(\sqrt{r_{1}}-\sqrt{r_{0}})^{2}}{\gamma}N_{S}\right]  ~,
\label{Fid_doubleCHECK}%
\end{equation}
where%
\begin{equation}
\gamma=1+(2-r_{0}-r_{1})N_{B}~, \label{gamma_ATT}%
\end{equation}
and
\begin{equation}
\theta=4N_{B}^{2}\prod\limits_{i=0,1}(1-r_{i})[1+(1-r_{i})N_{B}]~.
\label{theta_ATT}%
\end{equation}

Notice that the lower-bound $\mathcal{C}(M,N_{S})$ depends on the signal
profile $\{M,N_{S}\}$, but not from the number $L$ of idlers and the explicit
P-representation describing the classical state of the transmitter. As a
consequence, all the classical transmitters with the same signal profile
$\{M,N_{S}\}$\ are lower-bounded by $\mathcal{C}(M,N_{S})$. The average
information $J_{C}$ which is decoded from the memory cell is upper-bounded by
the quantity%
\begin{equation}
J_{class}:=1-H[\mathcal{C}(M,N_{S})]~.
\end{equation}

\subsubsection{Quantum transmitter}

Now, let us consider a quantum transmitter with the same signal profile
$\{M,N_{S}\}$\ but possessing EPR correlations between signals and idlers (EPR
quantum transmitter). In this case, we have the same number of signals and
idlers ($M=L$), and the global state for the input systems $\{S,I\}$ is a
tensor product of $M$ identical two-mode squeezed vacuum states, i.e.,
\begin{equation}
\rho_{SI}=\left\vert \xi\right\rangle _{si}\left\langle \xi\right\vert
^{\otimes M}~, \label{inputEPRtran}%
\end{equation}
where the single-copy state $\left\vert \xi\right\rangle _{si}\left\langle
\xi\right\vert $ refers to a single pair of signal and idler modes
$\{s,i\}\in\{S,I\}$. Recall that a two-mode squeezed vacuum state $\left\vert
\xi\right\rangle _{si}\left\langle \xi\right\vert $ is a Gaussian state with
zero mean and CM%
\begin{equation}
\mathbf{V}_{si}=\left(
\begin{array}
[c]{cc}%
(2N_{S}+1)\mathbf{I} & 2\sqrt{N_{S}(N_{S}+1)}\mathbf{Z}\\
2\sqrt{N_{S}(N_{S}+1)}\mathbf{Z} & (2N_{S}+1)\mathbf{I}%
\end{array}
\right)  ~, \label{CM_si}%
\end{equation}
where $\mathbf{I}=\mathrm{diag}(1,1)$, $\mathbf{Z}=\mathrm{diag}(1,-1)$ and
the squeezing parameter $\xi$ is connected to the signal-energy by the
relation $N_{S}=\sinh^{2}\xi$. At the output of the cell, the conditional
state of the systems $\{R,I\}$ is given by%
\begin{equation}
\rho_{RI}(u)=\rho_{ri}(u)^{\otimes M}~,
\end{equation}
where%
\begin{equation}
\rho_{ri}(u)=\left(  \mathcal{R}_{u}\otimes\mathcal{I}\right)  (\left\vert
\xi\right\rangle _{si}\left\langle \xi\right\vert ) \label{Rho_ri}%
\end{equation}
is the single-copy output state, i.e., describing a single pair of reflected
and idler modes $\{r,i\}\in\{R,I\}$. In fact, since the memory cell
corresponds to a one-mode channel and the state of the transmitter to a tensor
product, the output state at the receiver is also a tensor product state. In
particular, it corresponds to $M$ identical copies of the two-mode state of
Eq.~(\ref{Rho_ri}). Then, the decoding of $u$ corresponds to the $M$-copy
discrimination between the two states $\rho_{ri}(0)$ and $\rho_{ri}(1)$. The
corresponding minimum error probability $P_{err}^{quant}$ can be upper-bounded
by the quantum Chernoff bound, i.e.,%
\begin{equation}
P_{err}^{quant}\leq\mathcal{Q}(M,N_{S}):=\frac{1}{2}\left[  Q(N_{S})\right]
^{M}. \label{Q_M_Ns}%
\end{equation}
where%
\begin{equation}
Q(N_{S}):=\inf_{s\in(0,1)}\mathrm{Tr}\left[  \rho_{ri}(0)^{s}\rho
_{ri}(1)^{1-s}\right]  ~.
\end{equation}
Notice that the single-copy state $\rho_{ri}(u)$ is a Gaussian state with zero
mean and CM%
\begin{equation}
\mathbf{V}_{ri}(u)=\left(
\begin{array}
[c]{cc}%
\lbrack r_{u}\mu+(1-r_{u})\beta]\mathbf{I} & \sqrt{r_{u}(\mu^{2}-1)}%
\mathbf{Z}\\
\sqrt{r_{u}(\mu^{2}-1)}\mathbf{Z} & \mu\mathbf{I}%
\end{array}
\right)  ~, \label{outputCM_r}%
\end{equation}
where%
\begin{equation}
\mu:=2N_{S}+1~,~\beta:=2N_{B}+1~.
\end{equation}
Since $\rho_{ri}(0)$ and $\rho_{ri}(1)$ are two-mode Gaussian states, we can
compute the quantum Chernoff bound $\mathcal{Q}(M,N_{S})$ by using the formula
of Ref.~\cite{MinkoPRA}, which exploits the symplectic decomposition of the
Gaussian states. It is important to note that the CM of Eq.~(\ref{outputCM_r})
is in a special form, for which we can easily provide analytical expressions
for both the symplectic spectrum and the diagonalizing symplectic matrix
$\mathbf{S}$. In fact, let us set%
\begin{equation}
a:=r_{u}\mu+(1-r_{u})\beta~,~b:=\mu~,
\end{equation}
and%
\begin{equation}
c:=\sqrt{r_{u}(\mu^{2}-1)}\geq0~,
\end{equation}
so that the CM\ takes the special normal form%
\begin{equation}
\mathbf{V}=\left(
\begin{array}
[c]{cc}%
a\mathbf{I} & c\mathbf{Z}\\
c\mathbf{Z} & b\mathbf{I}%
\end{array}
\right)  ~. \label{V_simple}%
\end{equation}
The corresponding symplectic eigenvalues are given by~\cite{RMP}%
\begin{equation}
\nu_{1}=\frac{1}{2}\left(  \sqrt{y}+a-b\right)  ~, \label{Spectrum_simple1}%
\end{equation}
and%
\begin{equation}
\nu_{2}=\frac{1}{2}\left(  \sqrt{y}+b-a\right)  ~, \label{Spectrum_simple2}%
\end{equation}
where $y:=(a+b)^{2}-4c^{2}\geq4$. Thus, the Williamson form of $\mathbf{V}$ is
given by
\begin{gather}
\mathbf{W}=\nu_{1}\mathbf{I}\oplus\nu_{2}\mathbf{I}\nonumber\\
=\frac{1}{2}\left(
\begin{array}
[c]{cc}%
(\sqrt{y}+a-b)\mathbf{I} & \\
& (\sqrt{y}+b-a)\mathbf{I}%
\end{array}
\right)  .
\end{gather}
The symplectic matrix $\mathbf{S}$ which realizes the symplectic decomposition
$\mathbf{V}=\mathbf{SWS}^{T}$ is given by the formula~\cite{RMP}%
\begin{equation}
\mathbf{S}=\left(
\begin{array}
[c]{cc}%
x_{+}\mathbf{I} & x_{-}\mathbf{Z}\\
x_{-}\mathbf{Z} & x_{+}\mathbf{I}%
\end{array}
\right)  ~, \label{Symplectic_expression}%
\end{equation}
where%
\begin{equation}
x_{\pm}:=\sqrt{\frac{a+b\pm\sqrt{y}}{2\sqrt{y}}}\geq0~. \label{Terms_S}%
\end{equation}
Now, by expliciting the symplectic diagonalization for the two possible cases
$u=0$ and $u=1$, we have%
\begin{equation}
\mathbf{V}_{ri}(u)=\mathbf{S}(u)~\left[  \nu_{1}(u)\mathbf{I}\oplus\nu
_{2}(u)\mathbf{I}\right]  ~\mathbf{S}(u)^{T}~. \label{Symp_DIAG_k}%
\end{equation}
Using this decomposition, we can compute the quantum Chernoff bound by means
of the formula of Ref.~\cite{MinkoPRA}. Unfortunately, the analytical
expression is cumbersome, but we can easily derive numerical values for every
choice of the parameters.

It is clear that, from the upper bound $P_{err}^{quant}\leq\mathcal{Q}%
(M,N_{S})$, we can derive a lower bound for the average information $J_{Q}$
which is decoded via this quantum transmitter. This lower bound is simply
given by
\begin{equation}
J_{quant}:=1-H[\mathcal{Q}(M,N_{S})]~.
\end{equation}

\subsubsection{Comparison}

In order to compare quantum and classical reading, we fix the basic parameters
of the model $\{M,N_{S},r_{0},r_{1},N_{B}\}$ and we consider the difference%
\begin{equation}
G:=J_{quant}-J_{class}~,
\end{equation}
that we have called \textquotedblleft information gain\textquotedblright. It
is trivial to check that%
\begin{equation}
G\leq J_{Q}-J_{C}~.
\end{equation}
In other words, $G$ is a lowerbound for the average information which is
gained by using the EPR quantum transmitter instead of any classical
transmitter. A positive gain ($G>0$) is a sufficient condition for the
superiority of the quantum reading ($J_{Q}>J_{C}$). In general, this quantity
is a function of all the basic parameters of the model, i.e., $G=G(M,N_{S}%
,r_{0},r_{1},N_{B})$. Numerically, we can find signal profiles $\{M,N_{S}\}$,
classical memories $\{r_{0},r_{1}\}$, and thermal baths $N_{B}$, for which we
have the quantum effect $G>0$. Some of these values are shown by the table in
the main text. As explained in the main text, we can also resort to the
further lowerbound $G^{\ast}\leq G$, which is defined by using the quantum
Battacharyya bound instead of the quantum Chernoff bound. By exploiting
$G^{\ast}$, we can plot Figs.~\ref{FIGab}, \ref{FIGcd}, and~\ref{FIGef}.

\subsection{Ideal memories\label{AppIDEAL}}

Quantum reading is generally more powerful when the land-reflectivity is
sufficiently high (i.e., $r_{1}\gtrsim0.8$). For this reason, it is important
to analyze the scenario in the limit of ideal land-reflectivity ($r_{1}=1$),
defining the so-called \textquotedblleft ideal memories\textquotedblright. In
the presence of an ideal memory, one of the two possible outputs of the cell
is just the input state, i.e., we have%
\begin{equation}
\rho_{RI}(1)=\rho_{SI}~.
\end{equation}
Clearly, this fact leads to a simplification of the calculus. In the case of
an EPR quantum transmitter, the input state is pure and given by
Eq.~(\ref{inputEPRtran}). As a consequence, we have%
\begin{equation}
\rho_{RI}(1)=\rho_{ri}(1)^{\otimes M}=\left\vert \xi\right\rangle
_{si}\left\langle \xi\right\vert ^{\otimes M}~,
\end{equation}
i.e., one of the two output states is pure. As a consequence, the quantum
Chernoff bound can be reduced to the computation of the quantum fidelity. In
fact, we have~\cite{Gae}%
\begin{align}
Q(N_{S})  &  :=\inf_{s\in(0,1)}\mathrm{Tr}\left[  \rho_{ri}(0)^{s}\rho
_{ri}(1)^{1-s}\right] \nonumber\\
&  =\inf_{s\in(0,1)}\mathrm{Tr}\left[  \rho_{ri}(0)^{s}\left\vert
\xi\right\rangle _{si}\left\langle \xi\right\vert ^{1-s}\right] \nonumber\\
&  =\lim_{s\rightarrow1^{-}}\mathrm{Tr}\left[  \rho_{ri}(0)^{s}\left\vert
\xi\right\rangle _{si}\left\langle \xi\right\vert ^{1-s}\right] \nonumber\\
&  =F[\rho_{ri}(0),\left\vert \xi\right\rangle _{si}\left\langle
\xi\right\vert ]~,
\end{align}
where the fidelity $F[\rho_{ri}(0),\left\vert \xi\right\rangle _{si}%
\left\langle \xi\right\vert ]$ is between a mixed two-mode Gaussian state
$\rho_{ri}(0)$ with CM given in Eq.~(\ref{outputCM_r}) and a pure two-mode
Gaussian state $\left\vert \xi\right\rangle _{si}\left\langle \xi\right\vert $
with CM given in Eq.~(\ref{CM_si}). Then, we can apply the formula of
Ref.~\cite{Gae} for the quantum fidelity between multimode Gaussian states. We
achieve%
\begin{align}
Q(N_{S})  &  =F[\rho_{ri}(0),\left\vert \xi\right\rangle _{si}\left\langle
\xi\right\vert ]=\nonumber\\
&  =\frac{1}{[1+(1-\sqrt{r_{0}})N_{S}]^{2}+N_{B}(2N_{S}+1)(1-r_{0})}~,
\end{align}
which is the result given in the main text.

In the case of a classical transmitter, we just have to consider the lower
bound of Eq.~(\ref{CB_cread}) where now we set $r_{1}=1$ in the expression of
the fidelity given in Eq.~(\ref{Fid_doubleCHECK}). One can easily check that
the resulting fidelity takes the analytical form given in Eq.~(\ref{fidTEXT}).

\section{Technical proofs\label{AppTECH}}

Here, we explicitly prove the asymptotic expansions which have been presented
in the main text and used for the analysis of the ideal memories.

\subsection{General asymptote $(r_{0}=1)$\label{app1}}

According to Fig.~\ref{PRLmin}, the critical number $M^{(N_{S})}(r_{0})$
diverges for $r_{0}\rightarrow1$. Let us analyze the behavior of $G$ around
the singular point $r_{0}=1$, by setting $r_{0}=1-\varepsilon$ and expanding
$G$ for $\varepsilon\rightarrow0^{+}$. It is easy to check that, for every
$N_{B}$, we have $G>0$ if and only if
\begin{equation}
M>[4N_{S}(2N_{S}+1)\varepsilon]^{-1}~.
\end{equation}
In particular, in the absence of thermal noise ($N_{B}=0$), we have%
\begin{equation}
G=\frac{MN_{S}(4MN_{S}-1)\varepsilon^{2}}{8\ln2}+O(\varepsilon^{3})~,
\label{G_expans}%
\end{equation}
which is positive if and only if $M>(4N_{S})^{-1}$.

These conditions are easy to prove. In fact, note that $G>0$ if and only if
\begin{equation}
\Delta:=\mathcal{Q}(M,N_{S})-\mathcal{C}(M,N_{S})<0~.
\end{equation}
Thus, let us expand $\Delta=\Delta(M,N_{S},N_{B},1-\varepsilon)$\ at the first
order in $\varepsilon$. For a given $N_{B}>0$, we have
\begin{align}
\Delta &  =\frac{1}{2}\left[  \left(  MN_{B}\varepsilon\right)  ^{1/2}%
-M(N_{B}+N_{S}+2N_{B}N_{S})\varepsilon\right] \nonumber\\
&  +O(\varepsilon^{3/2})~,
\end{align}
which is negative if and only if%
\begin{equation}
M>\frac{N_{B}}{(N_{B}+N_{S}+2N_{B}N_{S})^{2}\varepsilon}:=\kappa(N_{B})~.
\end{equation}
Notice that $\kappa(N_{B})$ is maximum for
\begin{equation}
N_{B}^{\ast}=N_{S}(1+2N_{S})^{-1}~.
\end{equation}
Then, for every $N_{B}>0$, we have $\Delta<0$ if and only if%
\begin{equation}
M>\kappa(N_{B}^{\ast})=\frac{1}{4N_{S}(2N_{S}+1)\varepsilon}~. \label{Bound_M}%
\end{equation}
Now, let us consider the particular case of $N_{B}=0$. In this case, we have
the first-order expansion
\begin{equation}
\Delta=\left(  MN_{S}\right)  ^{1/2}[1-2\left(  MN_{S}\right)  ^{1/2}%
]\varepsilon/4+O(\varepsilon^{2})~,
\end{equation}
or, equivalently, the second-order expansion of $G$ given in
Eq.~(\ref{G_expans}). It is clear that $\Delta<0$, i.e., $G>0$, when
$M>1/4N_{S}$. However, this condition is less restrictive than the one in
Eq.~(\ref{Bound_M}) which, therefore, can be extended to every $N_{B}\geq0$.

\subsection{High-energy asymptote $(r_{0}=0)$\label{app2}}

\begin{figure}[ptbh]
\vspace{-0.0cm}
\par
\begin{center}
\includegraphics[width=0.25\textwidth] {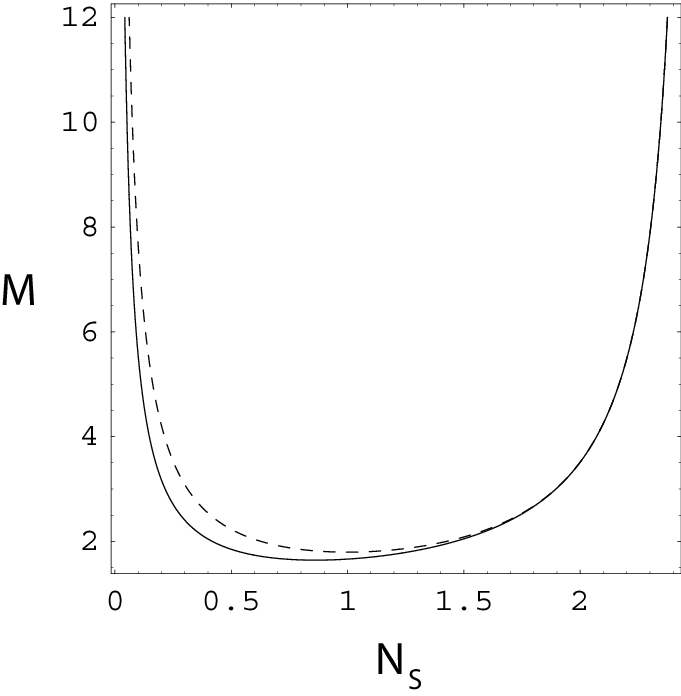}
\end{center}
\par
\vspace{-0.7cm}\caption{Minimum number of signals $M$ versus energy $N_{S}$.
The solid curve represents $M^{(N_{S})}(0)$ while the dashed curve represents
$\tilde{M}$. Notice that the minimum number of signals is actually given by
$\left\lceil M\right\rceil $ where $\left\lceil \cdots\right\rceil $ is the
ceiling function.}%
\label{ApproxPIC}%
\end{figure}

Let us analyze the behavior of $M^{(N_{S})}(r_{0})$ for $N_{S}\geq1$ and
$r_{0}=0$. One can check that, for $N_{S}\geq1$, the greatest value of
$M^{(N_{S})}(0)$\ occurs when $N_{B}=0$. In this case, i.e., for $r_{0}%
=N_{B}=0$ and $r_{1}=1$, we have
\begin{equation}
\mathcal{Q}(M,N_{S})=\frac{\left(  1+N_{S}\right)  ^{-2M}}{2}~,
\end{equation}
and%
\begin{equation}
\mathcal{C}(M,N_{S})=\frac{1-\sqrt{1-e^{-MN_{S}}}}{2}\overset{M\gg
1}{\longrightarrow}\frac{e^{-MN_{S}}}{4}:=\mathcal{C}^{\infty}.
\end{equation}
Let us consider the critical value $M^{(N_{S})}(0)$ of $M$ such that
$G(M,N_{S})=0$, which is equivalent to $\mathcal{Q}=\mathcal{C}$. We also
consider the value $\tilde{M}$ such that $\mathcal{Q}=\mathcal{C}^{\infty}$.
\ We find that $M^{(N_{S})}(0)\simeq\tilde{M}$ with very good approximation
when $N_{S}\geq1$ (see Fig.~\ref{ApproxPIC}). Then, for every $N_{S}\geq1$, we
can set
\begin{equation}
M^{(N_{S})}(0)\simeq\tilde{M}=(\ln2)\left[  2\ln(1+N_{S})-N_{S}\right]
^{-1}~.
\end{equation}
The latter quantity becomes infinite for $2\ln(1+N_{S})=N_{S}$, i.e., for
$N_{S}\gtrsim2.51$ photons.

\bigskip

\begin{thebibliography}{99}                                                                                               %


\bibitem {NielsenBook}M. A. Nielsen and I. L. Chuang, \textit{Quantum
Computation and Quantum Information }(Cambridge University Press, Cambridge, 2000).

\bibitem {BraREV}S. L. Braunstein and A. K. Pati, \textit{Quantum Information
Theory with Continuous Variables}, (Kluwer Academic, Dordrecht, 2003).

\bibitem {BraREV2}S. L. Braunstein and P. van Loock, Rev. Mod. Phys.
\textbf{77}, 513 (2005).

\bibitem {GaussSTATES}A. Ferraro, S. Olivares, and M. Paris, \textit{Gaussian
states in quantum information}, ISBN 88-7088-483-X (Biliopolis, Napoli, 2005).

\bibitem {GaussSTATES2}J. Eisert, and M.B. Plenio, Int. J. Quant. Inf.
\textbf{1}, 479 (2003).

\bibitem {RMP}C. Weedbrook, S. Pirandola, R. Garc\'{\i}a-Patr\'{o}n, N. J.
Cerf, T. C. Ralph, J. H. Shapiro, and S. Lloyd, Rev. Mod. Phys. \textbf{84},
621 (2012).

\bibitem {CVtelepo}A. Furusawa \textit{et al.}, Science \textbf{282}, 706 (1998).

\bibitem {Bra98}S. L. Braunstein, and H. J. Kimble, Phys. Rev. Lett.
\textbf{80}, 869 (1998).

\bibitem {RalphTELE}T. C. Ralph, Optics Letters \textbf{24}, 348 (1999).

\bibitem {PirTeleOPMecc}S. Pirandola, S. Mancini, D. Vitali, and P. Tombesi,
Phys. Rev. A \textbf{68}, 062317 (2003).

\bibitem {Barlett2003}S. D. Bartlett and W. J. Munro, Phys. Rev. Lett.
\textbf{90}, 117901 (2003).

\bibitem {Sherson}J. Sherson, H. Krauter, R. K. Olsson, B. Julsgaard, K.
Hammerer, I. Cirac, and E. S. Polzik, Nature \textbf{443}, 557 (2009).

\bibitem {TeleNET}P. van Loock, and S. L. Braunstein, Phys. Rev. Lett.
\textbf{84}, 3482 (2000).

\bibitem {teleREV}S. Pirandola and S. Mancini, Laser Physics \textbf{16}, 1418 (2006).

\bibitem {teleJMO}S. Pirandola, S. Mancini, D. Vitali, and P. Tombesi, J. Mod.
Opt. \textbf{51}, 901 (2004).

\bibitem {Pirgames}S. Pirandola, Int. J. Quant. Inf. \textbf{3}, 239 (2005).

\bibitem {Pir2005}S. Pirandola, S. Mancini, and D. Vitali, Phys. Rev. A
\textbf{71}, 042326 (2005).

\bibitem {Entswap}P. van Loock and S. L. Braunstein, Phys. Rev. A \textbf{61},
010302(R) (1999).

\bibitem {Entswap2}N. Takei, H. Yonezawa, T. Aoki, and A. Furusawa, Phys. Rev.
Lett. \textbf{94}, 220502 (2005).

\bibitem {PirENTswap}S. Pirandola, D. Vitali, P. Tombesi, and S. Lloyd, Phys.
Rev. Lett. \textbf{97}, 150403 (2006).

\bibitem {QKD0}N. J. Cerf, M. Levy, and G. van Assche, Phys. Rev. A
\textbf{63}, 052311 (2001).

\bibitem {QKD1}F. Grosshans, and P. Grangier, Phys. Rev. Lett. \textbf{88},
057902 (2002).

\bibitem {Weed}C. Weedbrook, A. M. Lance, W. P. Bowen, T. Symul, T. C. Ralph,
and P. K. Lam, Phys. Rev. Lett.\textbf{\ 93}, 170504 (2004).

\bibitem {Weed2}C. Weedbrook, C., A. M. Lance, W. P. Bowen, T. Symul, T. C.
Ralph, and P. K. Lam, Phys. Rev. A \textbf{73}, 022316 (2006).

\bibitem {CharacATT}S. Pirandola, S. L. Braunstein, and S. Lloyd, Phys. Rev.
Lett. \textbf{101}, 200504 (2008).

\bibitem {Chris}C. Weedbrook, S. Pirandola, S. Lloyd, T. C. Ralph, Phys. Rev.
Lett. \textbf{105}, 110501 (2010).

\bibitem {PirNATURE}S. Pirandola, S. Mancini, S. Lloyd, and S. L. Braunstein,
Nature Physics \textbf{4}, 726 (2008).

\bibitem {Devetak}I. Devetak, IEEE Trans. Inf. Theory \textbf{5}1, 44 (2005).

\bibitem {PirSKcapacity}S. Pirandola, R. Garc\'{\i}a-Patr\'{o}n, S. L.
Braunstein, and S. Lloyd, Phys. Rev. Lett. \textbf{102}, 050503 (2009).

\bibitem {QKDreview}V. Scarani, H. Bechmann-Pasquinucci, N. J. Cerf, M. Dusek,
N. Lutkenhaus, and M. Peev, Rev. Mod. Phys. \textbf{81}, 1301 (2009).

\bibitem {Qcomp1}S. Lloyd and S. L. Braunstein, Phys. Rev. Lett. \textbf{82},
1784 (1999).

\bibitem {Qcomp2}D. Gottesman, A. Kitaev, and J. Preskill, Phys. Rev. A
\textbf{64}, 012310 (2001).

\bibitem {Qcomp2b}B. C. Travaglione and G. J. Milburn, Phys. Rev. A
\textbf{66}, 052322 (2002).

\bibitem {Qcomp2c}S. Pirandola, S. Mancini, D. Vitali, and P. Tombesi,
Europhys. Lett. \textbf{68}, 323 (2004).

\bibitem {Qcomp2d}S. Glancy and E. Knill, Phys. Rev. A \textbf{73}, 012325 (2006).

\bibitem {Qcomp2e}S. Pirandola, S. Mancini, D. Vitali, and P. Tombesi, J.
Phys. B: At. Mol. Opt. Phys. \textbf{39}, 997 (2006).

\bibitem {Qcomp2f}S. Pirandola, S. Mancini, D. Vitali, and P. Tombesi, Eur.
Phys. J. D \textbf{37}, 283-290 (2006).

\bibitem {QcREF}A. P. Lund, T. C. Ralph, and H. L. Haselgrove, Phys. Rev.
Lett. \textbf{100}, 030503 (2008).

\bibitem {QcREF2}S. Sefi and P. van Loock, Phys. Rev. Lett. \textbf{107},
170501 (2011).

\bibitem {Qcomp3}R. Raussendorf, and H. J. Briegel, Phys. Rev. Lett.
\textbf{86}, 5188 (2001).

\bibitem {Qcomp5}N. C. Menicucci, P. van Loock, M. Gu, C. Weedbrook, T. C.
Ralph, and M. A. Nielsen, Phys. Rev. Lett. \textbf{97}, 110501 (2006).

\bibitem {Qcomp6}J. Zhang and S. L. Braunstein, Phys. Rev. A \textbf{73},
032318 (2006).

\bibitem {Qcomp7}S. T. Flammia, N. C. Menicucci, and O. Pfister, J. Phys. B
\textbf{42}, 114009 (2009).

\bibitem {clusterREF}N. C. Menicucci, X. Ma, and T. C. Ralph, Phys. Rev. Lett.
\textbf{104}, 250503 (2010).

\bibitem {clusterREF2}L. Aolita, A. J. Roncaglia, A. Ferraro, and A.
Ac\'{\i}n, Phys. Rev. Lett. \textbf{106}, 090501 (2011).

\bibitem {EPRpaper}A. Einstein, B. Podolsky, and N. Rosen, Phys. Rev.
\textbf{47}, 777 (1935).

\bibitem {NOTEshot}For two bosonic modes, $A$ and $B$, with quadratures
$\hat{q}_{A}$, $\hat{p}_{A}$, $\hat{q}_{B}$ and $\hat{p}_{B}$, one can define
the two operators $\hat{q}_{-}:=(\hat{q}_{A}-\hat{q}_{B})/\sqrt{2}$ (relative
position)\ and $\hat{p}_{+}:=(\hat{p}_{A}+\hat{p}_{B})/\sqrt{2}$ (total
momentum). Then, the system has EPR correlations (in these operators) if
$V(\hat{q}_{-})+V(\hat{p}_{+})<2\nu_{0}$, where $V(\cdot)$ is the variance,
and $\nu_{0}$ is the standard quantum limit ($\nu_{0}=1$ in this paper.) A
bosonic system with EPR correlations is entangled, but the contrary is not
necessarily true. The EPR\ correlations represent the most typical kind of
continuous variable entanglement, and are usually generated by parametric down conversion.

\bibitem {Prepres}E. C. G. Sudarshan, Phys. Rev. Lett. \textbf{10}, 277 (1963).

\bibitem {Prepres2}R. J. Glauber, Phys. Rev. \textbf{131}, 2766 (1963).

\bibitem {QIll1}S.-H. Tan, B. I. Erkmen, V. Giovannetti, S. Guha, S. Lloyd, L.
Maccone, S. Pirandola, and J. H. Shapiro, Phys. Rev. Lett. \textbf{101},
253601 (2008).

\bibitem {QIll2}S. Lloyd, Science \textbf{321}, 1463 (2008).

\bibitem {QIll3}J. H. Shapiro and Seth Lloyd, New J. Phys. \textbf{11}, 063045 (2009).

\bibitem {Guha}S. Guha and B. Erkmen, Phys. Rev. A \textbf{80}, 052310 (2009).

\bibitem {Devi}A. R. Usha Devi and A. K. Rajagopal, Phys. Rev. A \textbf{79},
062320 (2009).

\bibitem {YuenNair}H. P. Yuen, and R. Nair, Phys. Rev. A \textbf{80}, 023816 (2009).

\bibitem {QreadingPRL}S. Pirandola, Phys. Rev. Lett. \textbf{106}, 090504 (2011).

\bibitem {Nair11}R.~Nair, Phys. Rev. A \textbf{84}, 032312 (2011).

\bibitem {Hirota11}O.~Hirota, \textquotedblleft Error free Quantum Reading by
quasi Bell state of entangled coherent states,\textquotedblright\ e-print
arXiv:1108.4163 (2011).

\bibitem {Bisio11}A.~Bisio, M.~Dall'Arno, and G. M.~D'Ariano, Phys. Rev. A
\textbf{84}, 012310 (2011).

\bibitem {Arno11}M.~Dall'Arno, A.~Bisio, G. M.~D'Ariano, M.~Mikov\'{a},
M.~Je\v{z}ek, and M.~Du\v{s}ek, Phys. Rev. A \textbf{85}, 012308 (2012).

\bibitem {QreadCAP}S.~Pirandola, C.~Lupo, V.~Giovannetti, S.~Mancini, and S.
L.~Braunstein, New J. Phys. \textbf{13}, 113012 (2011).

\bibitem {Saikat2}S. Guha, Z. Dutton, R. Nair, J. Shapiro, and B. Yen,
\textquotedblleft Information Capacity of Quantum Reading,\textquotedblright%
\ in Laser Science, OSA Technical Digest (Optical Society of America, 2011),
paper LTuF2.

\bibitem {Saikat}S. Guha , S.-H. Tan, and M. M. Wilde, \textquotedblleft
Explicit capacity-achieving receivers for optical communication and quantum
reading,\textquotedblright\ e-print arXiv:1202.0518 (2012).

\bibitem {Cover}T. M. Cover and J. A. Thomas, \textit{Elements of Information
Theory} (Wiley, Hoboken, 2006).

\bibitem {Helstrom}C. W. Helstrom, \textit{Quantum detection and estimation
theory} (Academic Press, New York, 1976).

\bibitem {Fuchs}C. A. Fuchs and J. V. de Graaf, IEEE Trans. Inf. Theory
\textbf{45}, 1216 (1999).

\bibitem {FuchsThesis}C. Fuchs, PhD thesis (Univ. of New Mexico, Albuquerque, 1995).

\bibitem {QCbound}K. M. R. Audenaert \textit{et al.}, Phys. Rev. Lett.
\textbf{98}, 160501 (2007).

\bibitem {QCbound2}J. Calsamiglia \textit{et al.}, Phys. Rev. A \textbf{77},
032311 (2008).

\bibitem {QCbound3}M. Nussbaum and A. Szko\l a, Annals of Statistics
\textbf{37}, 1040-1057 (2009).

\bibitem {QCbound4}K. M. R. Audenaert, M. Nussbaum, A. Szko\l a, and F.
Verstraete, Comm. Math. Phys. \textbf{279}, 251--283 (2008).

\bibitem {MinkoPRA}S. Pirandola and S. Lloyd, Phys. Rev. A \textbf{78}, 012331 (2008).

\bibitem {Alex}S. Pirandola, A. Serafini, and S. Lloyd, Phys. Rev. A
\textbf{79}, 052327 (2009).

\bibitem {SIMONprinc}R. Simon, N. Mukunda, and B. Dutta, Phys. Rev. A
\textbf{49}, 1567 (1994).

\bibitem {Willy}J. Williamson, Am. J. Math. \textbf{58}, 141 (1936).

\bibitem {Gae}G. Spedalieri, C. Weedbrook, and S. Pirandola, J. Phys. A: Math. Theor. \textbf{46}, 025304 (2013).

\bibitem {fidG2}H. Nha, and H. J. Carmichael, Phys. Rev. A \textbf{71}, 032336 (2005).

\bibitem {fidG3}S. Olivares, M. G. A. Paris, and U. L. Andersen, Phys. Rev. A
\textbf{73}, 062330 (2006).

\bibitem {fidG4}H. Scutaru, J. Phys. A \textbf{31}, 3659 (1998).
\end{thebibliography}
\end{document}